\documentstyle[preprint,aps]{revtex}
\begin{document}

\draft
\begin{title}
{Theoretical Analysis of STM Experiments at Rutile
TiO$_2$ Surfaces}
\end{title}

\author{O. G\"{u}lseren, R. James, and D.W. Bullett}
\address{
School of Physics, University of Bath,
Claverton Down, Bath BA2 7AY, UK.
}

\date{2 October 1996}
\maketitle

\begin{abstract}

A first-principles atomic orbital-based electronic structure method is
used to investigate the low index surfaces of rutile Titanium Dioxide. 
The method is relatively cheap in computational terms, making it
attractive for the study of oxide surfaces, many of which undergo
large reconstructions, and may be governed by the presence of Oxygen
vacancy defects.

Calculated surface charge densities are presented for low-index
surfaces of TiO$_2$, and the relation of these results to experimental 
STM images is discussed. Atomic resolution images at these surfaces tend
to be produced at positive bias, probing states which largely consist
of unoccupied Ti 3$d$ bands, with a small contribution from O 2$p$. 
These experiments are particularly interesting since the O atoms tend 
to sit up to $1~$\AA\ above the Ti atoms, so providing a play-off 
between electronic and geometric structure in image formation.

\end{abstract}

\pacs{PACS numbers: 61.16.Ch, 68.35.Bs, 71.15.Fv, 73.20.At }


\section{Introduction}

There has been considerable interest in recent years in using scanning
probe microscopy to investigate the surfaces of transition
metal oxides. These materials, frequently used in technological
applications in which surface properties are crucial, are obvious
candidates for study by methods which potentially offer atomic resolution;
even a low-index ``clean'' surface may undergo considerable structural
relaxation, and may have its electronic and structural properties 
dominated by localised (O-vacancy) defects. For the semiconducting 
oxides, such as rutile TiO$_2$ and SnO$_2$, scanning {\it tunnelling} 
microscopy (STM) may be used. The aim of our work is to use computational
modelling to aid understanding of atomic resolution STM experiments at
oxide surfaces using TiO$_2$ (110) in particular as a model surface. At
least three groups have published atomic resolution images at this
surface~\cite{murray3,engel1,engel2,japan1,japan2,vanderbilt3}, 
whose structure appears to be heavily influenced by local stoichiometry.
Issues to be addressed include identification of the atoms responsible
for features seen in positive-bias images, and the extent to which we
expect STM to distinguish between the various structural models proposed
for the surface.

Unfortunately, the problems associated with modelling the full STM
experiment are considerable. The breaking of translational symmetry caused
even by a simple bulk-terminated surface make quantitative atom-based
calculations very compute expensive. The addition of a scanning tip,
not even periodic in the plane of the surface, and its associated electric
fields, compounds the problem, as does the extension
to systems with reconstructions, defects and adsorbates. 

Perhaps the most successful computational methods for modelling surfaces
are those based on {\it ab initio} plane-wave pseudopotential
calculations. These have been performed for low-index TiO$_2$
surfaces~\cite{vanderbilt3,gillan,vanderbilt1,vanderbilt2}. Their principal
drawback is that atomic Oxygen is particularly difficult to pseudise, due
to the lack of core $p$ orbitals. The same problem occurs for the first-row
transition elements, lacking a core $d$ orbital. Pseudopotential
calculations involving these atoms are therefore rather expensive.
We are engaged in a programme of research using less rigorous, but far
cheaper, computational methods to model transition metal oxides. 
The ability of our method to reproduce a reasonable account of the
physics of these surfaces has been investigated through a series of
benchmarking exercises, reported elsewhere~\cite{oguz}.
Though the results presented here are for relatively simple, isolated
surfaces, we feel they already make a contribution to the
understanding of STM experiments at these surfaces. Further,
we are confident that it will be possible to model 
realistic systems, including tips, on a moderate workstation.

The rest of the paper is organised as follows. Section 2 contains a
brief outline of the computational method employed. In Section 3 we present 
calculated densities of states (DOS) for bulk rutile TiO$_2$, from which 
it might be assumed that positive bias experiments only ever image Ti atoms, 
and computed local-DOS (equivalent to the ``perfect tip'' approximation 
to STM imaging~\cite{tersoff}) which show that this is not necessarily the
case. Our conclusions are drawn in Section 4.

\section{Method}

The calculations were carried out in a first-principles atomic-orbital
based scheme~\cite{dave1}. The basis set comprised the valence orbitals
for Ti (4$s$, 4$p$, and 3$d$) and O (2$s$ and 2$p$). The numerical atomic
orbitals were generated using a standard local-density approximation for
exchange and correlation, and the potential in the solid was calculated by
superposing neutral-atom charge densities. The Schr\"{o}dinger equation was
then solved for the electronic structure of TiO$_2$ slabs with a thickness
of 6--12 atomic layers (sufficient to prevent substantial interaction
between the two surfaces). Self-consistency was included only to the extent
that the energy of the Ti 3$d$ state was made consistent with that in a
neutral atom with the same $d$-occupancy; this leads to an occupancy of
1.85 $d$ electrons per Ti atom in bulk TiO$_2$, with small variations from
this value at the surface. The method has previously been used to study the
electronic structure of various TiO$_2$ surfaces~\cite{purton1,purton2}.
Although this approach is crude compared to fully self-consistent
calculations, the advantage of the method is its simplicity which allows
it to be easily applied to the local electronic structure in complicated
geometries such as defective TiO$_2$ surfaces.

\section{Results}

The calculated DOS of bulk rutile TiO$_2$ is presented in
figure~\ref{fig:dos}. The total DOS is resolved in terms of atomic Ti and
O contributions. TiO$_2$ is a direct-gap semiconducting material. Our
calculations yield a fundamental gap of 2.85~eV; the experimental value is
3.0~eV~\cite{pasc}. The character of the valence band is mainly O 2$p$,
with rather less Ti 3$d$. The unoccupied, conduction band is formed by
Ti 3$d$ orbitals with a small contribution from O 2$p$ states. 

STM experiments with atomic resolution at TiO$_2$ surfaces are achieved
with a positive tip
bias~\cite{murray3,engel1,engel2,japan1,japan2,murray1,murray2}, 
which means the tunnelling occurs into empty conduction band states.
Because of the mainly Ti 3$d$ character of these states, it is expected
from the electronic point of view that Ti atoms will be imaged. However,
at the unrelaxed (100) and (110) surfaces, the uppermost O atoms are
approximately 1\AA\ above the Ti atoms. Since the tails of the radial
wavefunctions are decaying exponentially, the contribution of O atoms
to the STM images may be appreciable because of this height difference.
Radial wavefunctions of the relevant O and Ti atomic orbitals are shown
in figure~\ref{fig:dos}(b). In (c) the wavefunctions are re-plotted to
give an approximate indication of the interplay between electronic
structure and geometry. The O wavefunctions are shifted 1\AA\ to
the right of those of Ti to take into account the height difference
and are scaled by 0.1 since the O contribution to the DOS is $\sim$10\%
of the Ti contribution for the conduction band states shown. This plot
would suggest that both O and Ti may contribute to constant height images,
with O dominating out to $\sim$3\AA\ and Ti further out.

This simple analysis agrees well with the computed constant height ($z$)
charge densities $\rho(x,y)$, shown in figure~\ref{fig:surf100} for the 
unrelaxed, stoichiometric TiO$_2$ (100)$1\times 1$ surface. 
The charge densities consist of the sum of all states at $\Gamma$ within
1.5eV of the conduction band edge. Assuming the Fermi level is pinned at
or close to the conduction band edge, this figure represents the
crudest, perfect delta-function tip, estimate of a constant-height
STM image at +1.5V bias. At the smallest height $z = 1$\AA\ 
(measured from the uppermost surface O) $\rho(x,y)$ shows a peak close
to the surface O atom position, as seen in figure~\ref{fig:surf100}(b).
At $z=2$\AA\ $\rho(x,y)$ is more uniform along the 001 direction. Further
away from the surface at $z=4$\AA\ a peak appears close to the Ti atom
positions (figure~\ref{fig:surf100}d). By 6\AA\ our calculations suggest
that lateral structure has all but disappeared. This may give a qualitative
explanation as to why the $1\times 1$ surface has never been imaged with
atomic resolution in STM.

Atomically resolved STM images have been reported both for the 
$1\times 1$ and $1\times 2$ phases of 
TiO$_2$ (110)~\cite{murray3,engel1,engel2,japan1,japan2}. 
One of the proposed models for the $1\times 2$ surface is a missing row
reconstruction, in which every second bridging O row is removed
(figure~\ref{fig:surf110}a). This reduction results in a surface state just
below the bottom of the conduction band. Computed constant-height charge
densities $\rho(x,y)$ for the unrelaxed $1\times 1$ and unrelaxed missing
row $1\times 2$ surfaces are shown in figure~\ref{fig:surf110} at heights
of 2\AA\ and 4\AA. Again, only states at $\Gamma$ within 1.5eV of the
conduction band edge have been included. Two surface units cells of the
$1\times 1$ surface are plotted for easy comparison with the $1\times 2$
surface. At 2\AA\ the charge comes mainly from a state having the character
of O 2$p$ and two Ti 3$d$ orbitals. The mixing of all these results in a
peak over the bridging O atom. Thus the difference between the $1\times
1$ (b) and $1\times 2$ (c) plots is merely that half the bridging
Oxygens are missing in the latter.

At $z=4$\AA\ the charge density has switched, as in the (100) case, 
to being associated in position with surface Ti atoms~\cite{murray4}, 
though there is little to suggest much resolution of individual atoms
within rows lying along the [001] direction. At the $1\times 1$ surface,
the rows are centred over the fivefold-coordinated Ti atoms which sit
halfway between the rows of bridging Oxygen atoms. At the $1\times 2$
surface, there is a single, much wider row, centred over the
fourfold-coordinated Ti atoms exposed on the removal of the O rows.
These results, including the relative register of the main features in
the two phases, are in broad agreement with experimental STM images of
TiO$_2$ (110).

\section{Conclusions}

We have presented the results of computationally cheap atomic-orbital
based calculations at low-index TiO$_2$ surfaces which show broad
agreement with published experimental STM images.
Despite the problems inherent to any attempt to model STM, we are
encouraged that even the simplest calculations of charge density at the
surfaces of TiO$_2$ yield results which make sense in terms of published
STM images and yield information, relating to orbital content of
observed features, which may be of use in interpretation.
The constant height charge densities $\rho(x,y)$ at TiO$_2$ surfaces are
affected by both geometrical structure, as the surface Oxygen atoms
sit $\sim$  1\AA~above the next plane, and electronic structure, in that
the conduction band is mainly formed by Ti 3d orbitals but mixing of
orbitals shift the peak positions. The next stage at the
TiO$_2$(110)$1\times 2$ surface is to investigate the sensitivity of the
computed charge densities (integrated over the surface Brillouin zone)
to the geometry of various proposed structures. The inclusion of a tip,
and consideration of other oxide surfaces are obvious extensions on
which we are currently working.

The authors would like to thank Geoff Thornton and Paul Murray for
useful discussions, and the UK EPSRC for financial support.


\clearpage

\begin{figure}
\caption{a)~Computed density of states for bulk rutile TiO$_2$.
b)~Radial wavefunctions, multiplied by radius and squared, of the 
self-consistent atomic basis states of neutral Titanium and Oxygen.
c)~Some of the same states replotted to show how the charge density
at a surface depends on the relative heights and occupancy of contributing
orbitals. See text for details.}
\label{fig:dos}
\end{figure}

\begin{figure}
\caption{a)~The unrelaxed stoichiometric (100) surface of TiO$_2$. Dark
circles represent Ti atoms. b)--d)~Constant height charge densities
including only conduction band states at $\Gamma$ within 1.5~eV of the
conduction band edge at heights of 1, 2 and 4\AA\ respectively above the
uppermost O atoms, which sit below the centre of each figure. In each case,
a single surface unit cell is shown. In all greyscale images white shows
higher charge.}
\label{fig:surf100}
\end{figure}

\begin{figure}
\caption{a)~The unrelaxed missing O-row model of TiO$_2$(110)
$1\times2$, in which alternate bridging O rows are removed from the
stoichiometric $1\times 1$ termination. b)~Constant height charge
density at 2\AA\ above the bridging O atoms for two surface units cells
of stoichiometric TiO$_2$(110) $1\times 1$. c)~Constant height charge
density 2\AA\ above the bridging O atoms for a single surface unit cell
of the $1\times 2$ surface shown in a). d) and e) are the same as b) and c)
respectively, except at a height of 4\AA. All $\rho(x,y)$ consist of the sum
of conduction band states at $\Gamma$ within 1.5~eV of the conduction band
edge. In b)--e) each corner of the figure sits above a bridging O atom.
In all greyscale images white shows higher charge.}
\label{fig:surf110}
\end{figure}

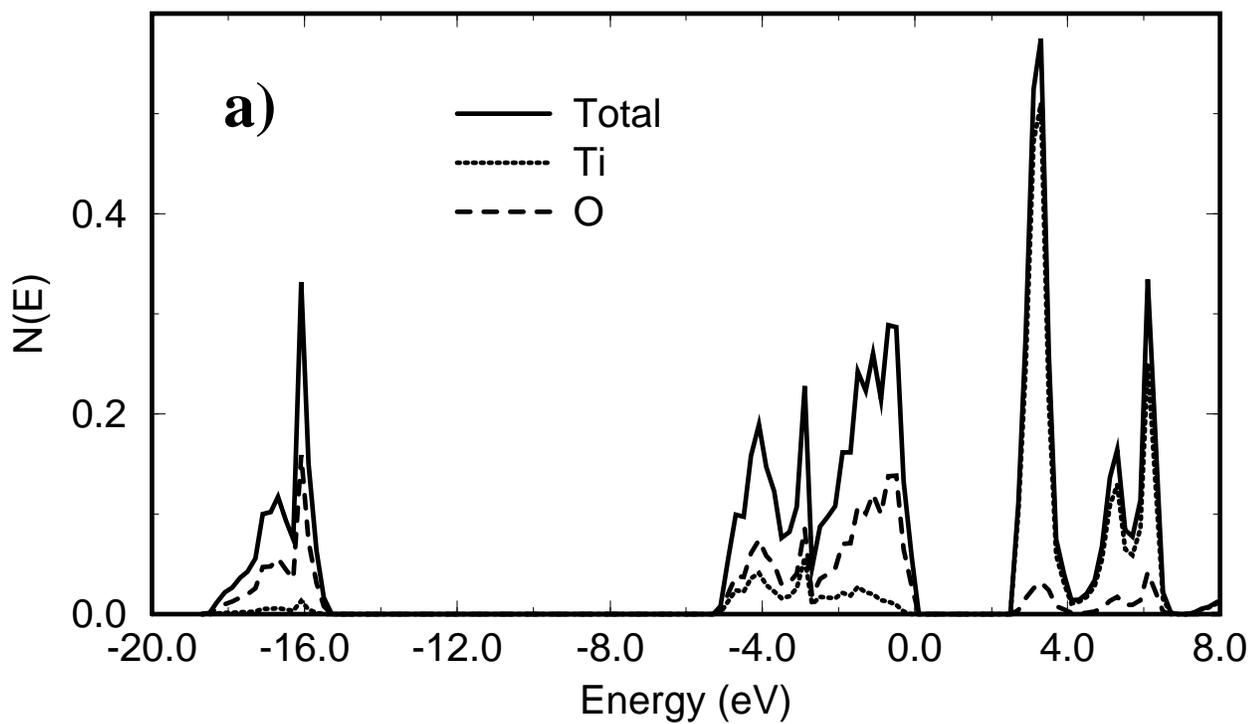
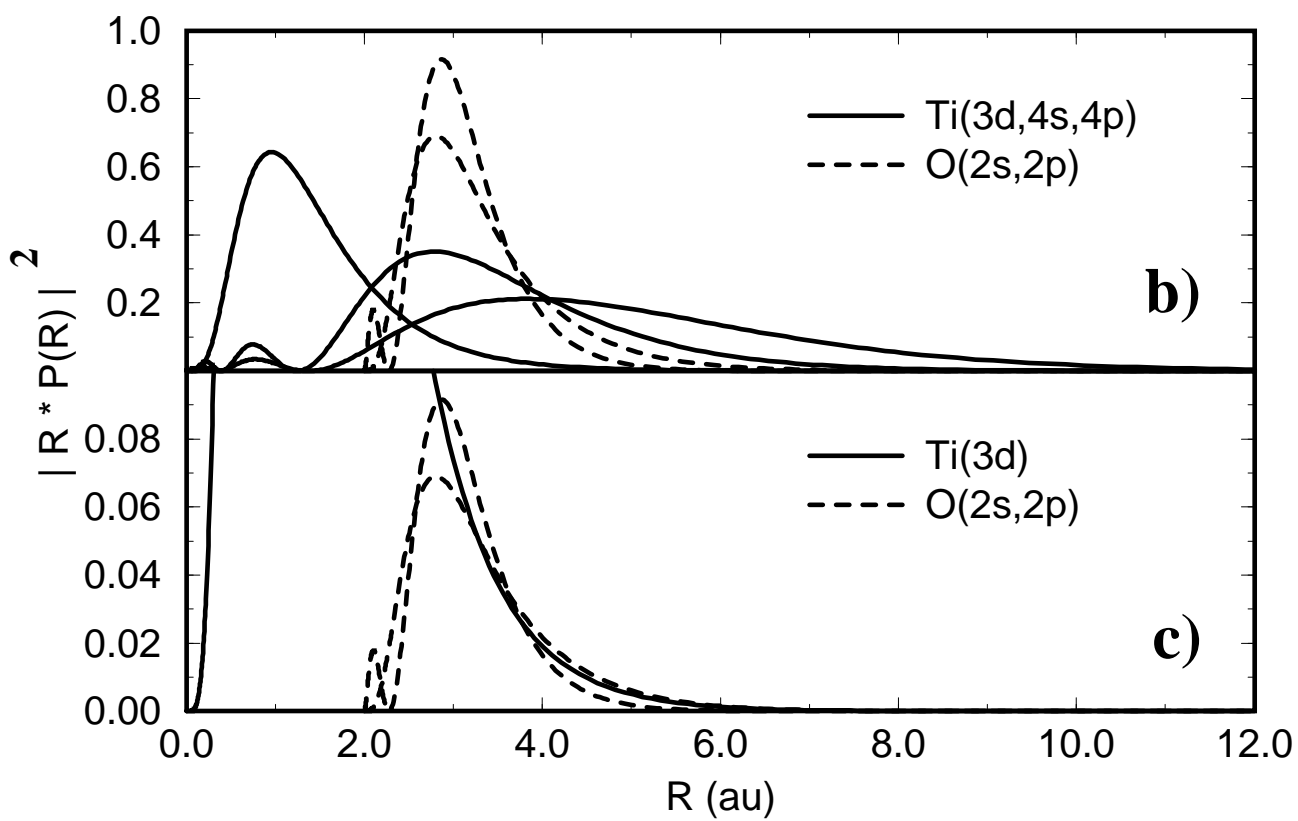
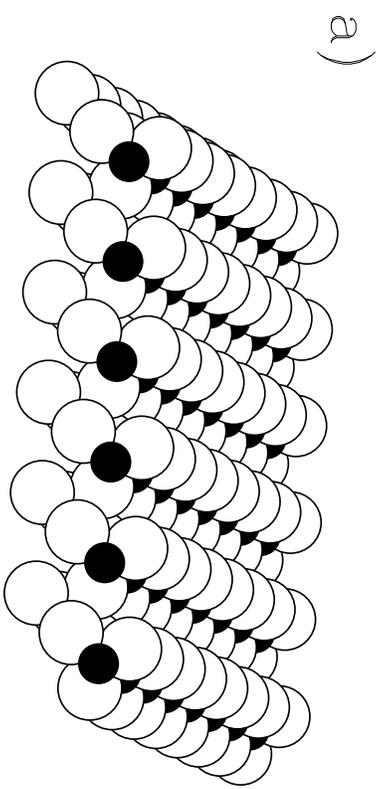
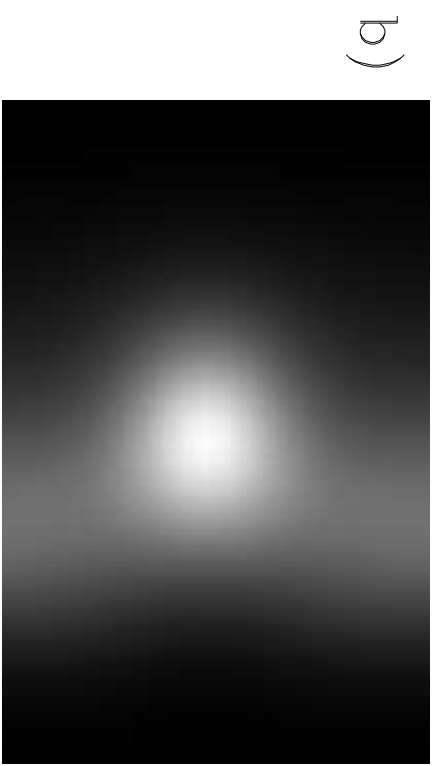
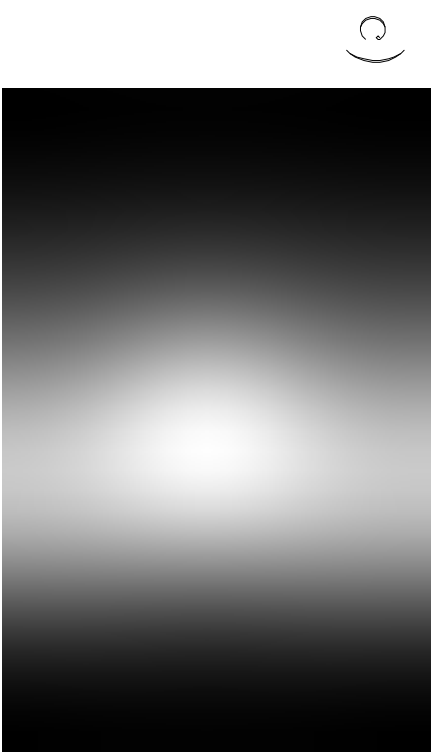
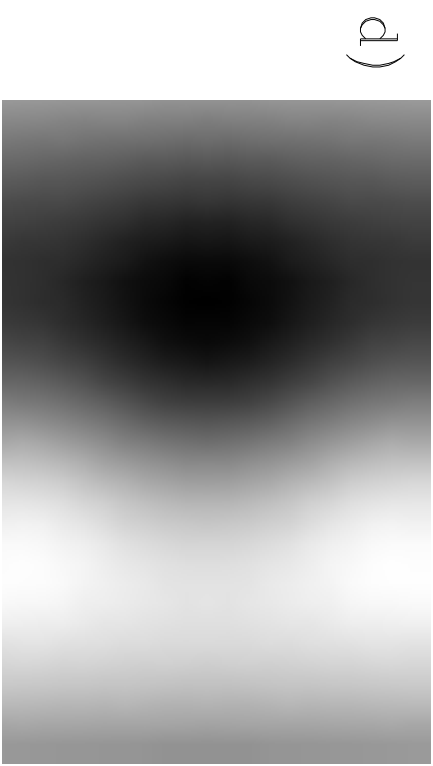
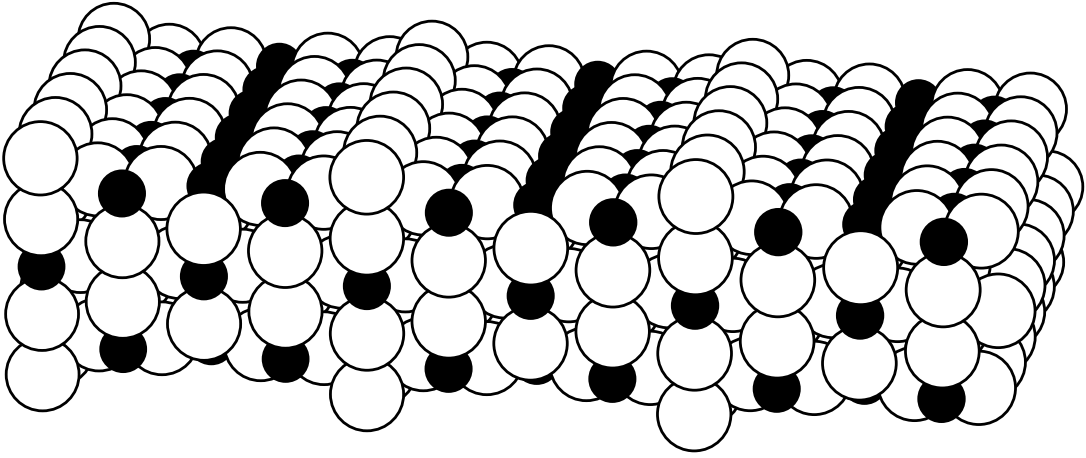
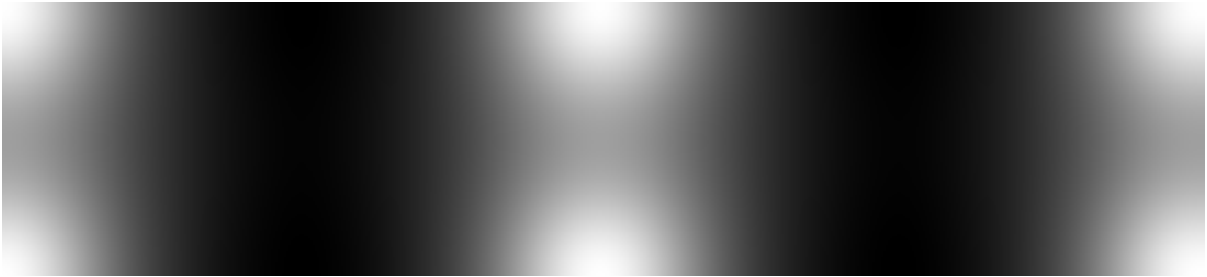
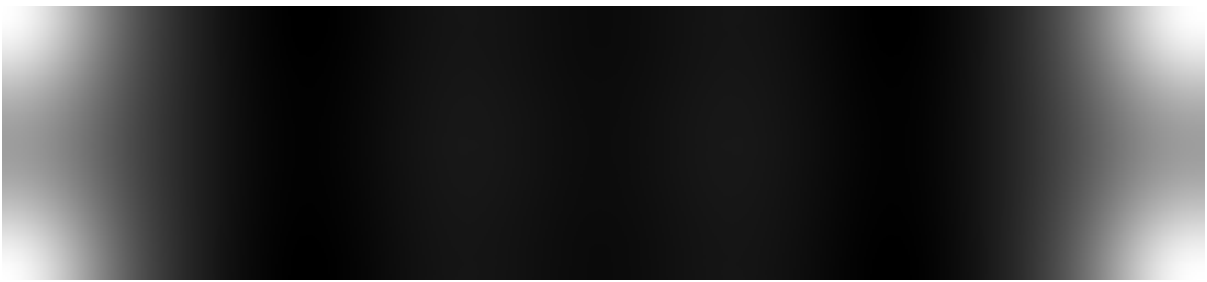
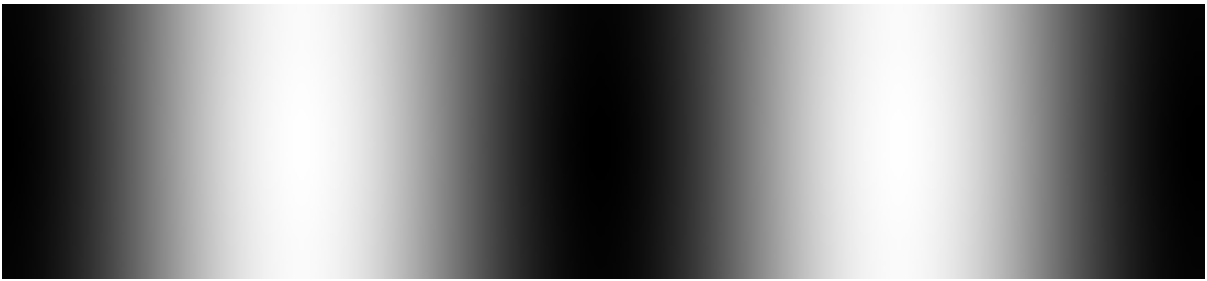
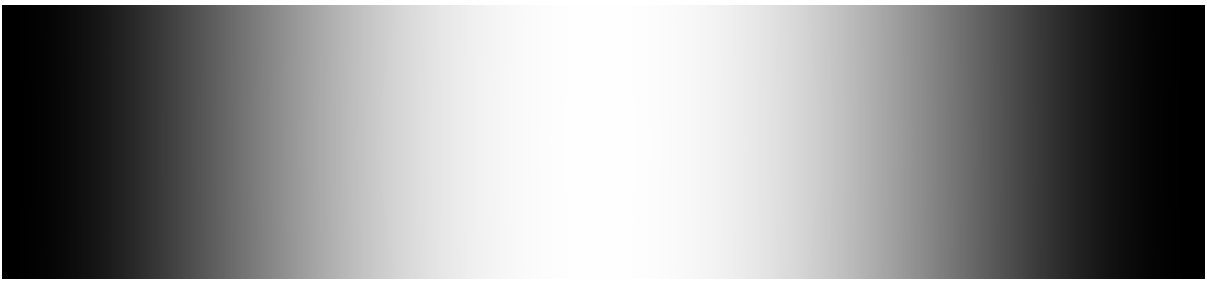
\end{document}